\documentclass[prl,amsmath,superscriptaddress,twocolumn,showpacs,showkeys]{revtex4}
\usepackage{bm}
\usepackage{amssymb}
\usepackage{graphicx}

\newcommand{\up}{\uparrow}
\newcommand{\down}{\downarrow}

\def\nab{{\mbox{\boldmath{$\nabla$}}}}

\begin{document}

\title{Limitations on the Ginzburg criterion for dirty superconductors}

\author{A. Aharony}
\email{aaharony@bgu.ac.il}
\altaffiliation{Also at Tel Aviv University.}

\affiliation{Department of Physics, Ben Gurion University, Beer
Sheva 84105, Israel}

\affiliation{ Ilse Katz Center for
Meso- and Nano-Scale Science and Technology, Ben Gurion
University, Beer Sheva 84105, Israel}

\author{O. Entin-Wohlman}
\altaffiliation{Also at Tel Aviv University.}
\affiliation{Department of Physics, Ben Gurion University, Beer
Sheva 84105, Israel}

\affiliation{ Ilse Katz Center for
Meso- and Nano-Scale Science and Technology, Ben Gurion
University, Beer Sheva 84105, Israel}

\author{H. Bary-Soroker}

\affiliation{Department of Physics, Ben Gurion University, Beer
Sheva 84105, Israel}

\author{Y. Imry}

\affiliation{Department of Condensed Matter Physics,  Weizmann
Institute of Science, Rehovot 76100, Israel}

\date{\today}

\begin{abstract}
The contributions of superconducting fluctuations to
the specific heat of dirty superconductors are calculated, including quantum and classical corrections to the `usual' leading Gaussian divergence.  These additional terms modify the Ginzburg criterion, which is based on equating these fluctuation-generated contributions to the mean-field discontinuity in the  specific heat, and set limits on its applicability for materials with a low transition temperature.

\end{abstract}

\pacs{74.40.-n,74.20.De,74.62.En}

\keywords{dirty superconductors, Ginzburg criterion, superconducting fluctuations, specific heat}

\maketitle

\section{Introduction}

The superconducting phase transition has been very well described by the Ginzburg-Landau (GL) theory, \cite{GL} which is equivalent to the BCS theory \cite {BCS} for temperatures $T$ close to the transition temperature $T^{}_c$. However, both theories use mean fields, and therefore miss the effects of critical fluctuations as $T$ approaches $T^{}_c$. \cite{Ginzburg}
A phenomenological way to estimate the region of validity of these mean-field theories is to calculate a certain effect of the fluctuations, and to require that this effect be small compared to the corresponding mean-field prediction. The relative temperature range for which the mean-field theory breaks down, $t^{}_G\equiv (T^{}_G-T^{}_c)/T^{}_c$, is called the ``Ginzburg region". \cite{Ginzburg}

Here we discuss dirty superconductors, which contain non-magnetic impurities. It is well known that such impurities reduce the superconducting coherence length, while leaving the transition temperature, as well as the density of states, largely unchanged. \cite{BCS,sasha} As a result, although the Ginzburg region is still rather small, it is significantly larger than for the pure superconductors. For $T>T^{}_c$, the superconducting order parameter vanishes, and the GL contribution to the specific heat vanishes. This specific heat has a discontinuity $\Delta C$ at $T^{}_c$. In contrast, the fluctuations in the order parameter
generate non-zero contributions $C^{}_{\rm fl}$ to the specific heat even above $T^{}_c$. One common way to estimate the Ginzburg region is then to require that outside of this regime one has $C^{}_{\rm fl}<\Delta C$. \cite{VARLAMOV}
As $T$ approaches $T^{}_c$, $C^{}_{\rm fl}$ diverges in $d<4$ dimensions as $|t|^{(d-4)/2}$, where $t\equiv \ln(T/T^{}_c)\approx (T-T^{}_c)/T^{}_c$. Keeping only this leading divergent term, one finds  $|t^{}_G| \sim (\Delta C/\Lambda^d)^{2/(d-4)}$, where $\Lambda$ is the momentum cutoff, which is the inverse of the relevant size of the fluctuations in space (to be discussed below).

The above result ignores corrections to the leading divergent term in $C^{}_{\rm fl}$. As we discuss below, such corrections arise both from quantum fluctuations and from corrections to the leading `static' contribution. The aim of the present paper is to discuss the effects of these corrections, which become crucial as $\Delta C$ becomes small.
In order to derive these corrections, it is important to obtain the full expression for the leading wave-vector and frequency dependent order-parameter correlation functions, and not just the `static' Ornstein-Zernike (OZ) \cite{OZ} approximation $\chi(q)\sim 1/(q^2+\xi^{-2})$ which is used in the
`standard' GL theory ($\xi$ is the coherence length). The derivation of this full expression is reviewed in Sec. II. Section III then presents the resulting contributions of the fluctuation to the specific heat, $C^{}_{\rm fl}$, including all the corrections, and Sec. IV discusses the consequences for the Ginzburg region. The results are summarized and discussed in Sec. V.

\section{The partition function}

\label{DETC}

We begin by reviewing the microscopic derivation of the free energy which
determines the superconducting fluctuations. The Hamiltonian  is
\begin{align}
{\cal H}=\int d{\bf r}{\cal H}({\bf r})\ ,
\end{align}
with
\begin{align}
{\cal H}({\bf r})&=\sum_{\sigma}\psi_{\sigma}^{\dagger}({\bf r})
{\cal H}^{}_{0}({\bf r})\psi^{}_{\sigma}({\bf r})\nonumber\\
&-V({\bf r})\psi^{\dagger}_{\up}({\bf
r})\psi^{\dagger}_{\down}({\bf r})\psi^{}_{\down}({\bf r})
\psi^{}_{\up}({\bf r})\ , \label{H}
\end{align}
where $\psi_{\sigma}^{\dagger}({\bf r})$ creates an electron with
spin $\sigma$ at ${\bf r}$. The interaction $V({\bf r})$ depends
on the spatial coordinate ${\bf r}$,
\begin{align}
V({\bf r})=\lambda({\bf r})/{\cal N}({\bf r})\ ,
\end{align}
where $\lambda({\bf r})$ is the effective local (dimensionless) electronic coupling, while
 ${\cal N}({\bf r})$ is the local density of states {\em
per unit volume and unit energy}.  The single-particle part of
the Hamiltonian (\ref{H}) reads
\begin{align}
{\cal H}^{}_{0}= -\nab^{2}/(2m)+u({\bf r})-\mu\
,\label{Hzero}
\end{align}
where $\mu$ is the chemical potential and
the disorder potential $u({\bf r})$ is modeled by point-like non-magnetic scatterers. \cite{AGD}

The following calculation has been given, for quasi-one dimensional rings, in Refs. \onlinecite{HAMUTAL2} and \onlinecite{we}. However, we repeat it here for the general $d$-dimensional case, in order to highlight the origin of the two Matsubara frequencies which generate the slow momentum and frequency variation of the quadratic coefficient in our effective GL theory. The quantum partition function ${\cal Z}$ is \cite{ALTLAND}
\begin{align}
{\cal Z}=\int {\cal D}\{\psi ({\bf r},\tau ),\overline{\psi}({\bf
r},\tau )\}\exp [-{\cal S}]\ ,\label{ZZ}
\end{align}
where the action ${\cal S}$ is
\begin{align}
{\cal S}=\int d{\bf r}\int_{0}^{\beta}d\tau \Bigl
(\sum_{\sigma}\overline{\psi}^{}_{\sigma} ({\bf r},\tau )
\frac{\partial}{\partial\tau} \psi^{}_{\sigma} ({\bf r},\tau
)+{\cal H}({\bf r},\tau )\Bigr )\ ,\label{S}
\end{align}
and $\beta=1/T$ (we use $\hbar=k^{}_B=1$). Here, the annihilation and creation  field
operators  in the Hamiltonian  (\ref{H}) ($\psi$ and $\psi^\dagger$) are replaced by the
 Grassmann variables $\psi ({\bf r},\tau )$ and
$\overline{\psi}({\bf r},\tau )$, respectively.

Applying the Hubbard-Stratonovich transformation to Eq.
(\ref{ZZ}), and integrating the fermionic part of the action, the
partition function is cast into the form \cite{ALTLAND}
\begin{align}
{\cal Z}&=\int {\cal D}\{\Delta ({\bf r},\tau ),\Delta^{\ast}_{}({\bf r},\tau )\}e^{-{\cal S}}\ ,\label{ZD}
\end{align}
with the action
\begin{align}
{\cal S}=\int d{\bf r}\int_{0}^{\beta} d\tau\frac{|\Delta ({\bf r},\tau )|^{2}}{V({\bf r})}-{\rm Tr}
\Bigl \{\ln\Bigl (\beta{\cal G}^{-1}\Bigr )\Bigr \}\ ,\label{SSO}
\end{align}
where  ${\cal G}^{-1}$ is the inverse
($2\times 2$ matrix) Green function at equal positions and imaginary times,
\begin{align}
{\cal G}^{-1}=\left [\begin{array}{cc}G^{-1}_{p}&\Delta \\ \Delta^{\ast}&G^{-1}_{h}
\end{array}\right ]\ ,
\end{align}
with
\begin{align}
G^{-1}_{p}&=-\partial^{}_{\tau}-{\cal H}^{}_0
\end{align}
being the particle inverse Green function, and
\begin{align}
G^{-1}_{h}&=
-\partial^{}_{\tau}+{\cal H}^{}_0
\end{align}
being the inverse Green function of the holes. 

The integration over the bosonic fields in Eq. (\ref{ZD}) is carried out using a stationary-phase
analysis  \cite{ALTLAND} of the action ${\cal S}$. At temperatures above the transition temperature,
this amounts to expanding the second term on the
right-hand side of Eq. (\ref{SSO})  to second order in $\Delta$ (the first-order contribution
 to the expansion being zero)
\begin{align}
&{\rm Tr}\{\ln (\beta {\cal G}^{-1})\}\Big |^{}_{\rm 2^{nd}}={\rm Tr}\Bigl \{\ln\beta\left [\begin{array}{cc}
G^{-1}_{p}&0 \\ 0&G^{-1}_{h}\end{array}\right ]\Bigr \}\nonumber\\
&+\int \frac{d{\bf r}d{\bf r}'}{\Omega^{2}}\int _{0}^{\beta}
\frac{d\tau d\tau '}{\beta^{2}}\Pi({\bf r},{\bf r}',\tau -\tau ' )\Delta ({\bf r}',\tau ')
\Delta^{\ast}_{}({\bf r},\tau )\ ,\label{EXPD}
\end{align}
where $\Omega $ denotes the volume of the system. 
The first term on the right-hand side of Eq. (\ref{EXPD})
gives the partition function of noninteracting electrons; the
second one represents the contribution of the superconducting
fluctuations to that function. Its calculation requires the
correlation
\begin{align}
&\Pi({\bf r},{\bf r}',\tau -\tau ')\equiv  \nonumber \\  &   - \langle G^{}_{p}({\bf r},{\bf r}',\tau -\tau ')
G^{}_{h}({\bf r}',{\bf r},\tau '-\tau )\rangle \ ,
\end{align}
where $\langle\ldots\rangle$ indicates averaging over the impurity configurations
(see Ref. \onlinecite{AGD}~ for details). For the general case, when the material contains several different regions (as e.g. in a double layer \cite{we}), this average depends separately on ${\bf r}$ and on ${\bf r}'$, and the calculation becomes difficult. However, for the purposes of the present discussion, it suffices to consider the simplest case, in which the material is homogeneous, and therefore the attractive interaction $V({\bf r})\equiv V$ as well as the density of states ${\cal N}$ are constant in space.  Averaging over the impurities restores homogeneity even in the dirty case, and the spatial dependence of $\Pi$
becomes a function of  ${\bf r-r}'$.  Hence,
\begin{align}
{\rm Tr}\{\ln (\beta {\cal G}^{-1})\}\Big |^{}_{\rm 2^{nd}}  = \sum_{\nu}\sum_{{\bf
q}}\Pi({\bf q},\nu)
|\Delta({\bf q},\nu)|^2\ ,\label{EXPS}
\end{align}
where
\begin{align}
\Pi({\bf q},\nu)=\sum_{{\bf p}^{}_1,{\bf
p}^{}_2}\sum_{\omega }& \langle G({\bf
p}^{}_1+{\bf q},{\bf p}^{}_2+{\bf q},\omega+\nu ) \nonumber \\  & \times
 G^{}_{}(-{\bf p}^{}_1,-{\bf
p}^{}_2,-\omega)\rangle\ ,\label{KXX}
\end{align}
and both Green functions are the particle ones, \cite{HAMUTAL2}
i.e., $G=G_{p}$.  We use the notations $\omega
\equiv\omega^{}_{n}=\pi T(2n+1)$ for the fermionic Matsubara
frequencies, and $\nu\equiv\nu^{}_{m}=2\pi Tm$ for the
bosonic ones ($n$ and $m$ are integers). Since the phonon-mediated electron-electron attractive interaction is
limited to energies within the Debye frequency $\omega^{}_D$ from the Fermi energy, both $|\omega|$ and $|\omega+\nu|$ are bound by $\omega^{}_D$.

Inserting these results into the expression for the action [see
Eq. (\ref{SSO})], the Gaussian fluctuation-induced partition
function, ${\cal Z}_{\rm fl,2}$, takes the form
\begin{align}
{\cal Z}^{}_{\rm fl,2}=\int {\cal D}\{\Delta ({\bf
q},\nu),\Delta^{\ast}_{} ({\bf q},\nu)\}e^{-{\cal S}^{}_2}\ ,\label{Zfl}
\end{align}
with
\begin{align}
{\cal S}^{}_2=\sum_{\bf q}\sum_{\nu} \Delta^{\ast}_{}
({\bf q},\nu)\Bigl (\frac{\beta\Omega}{V}-\Pi({\bf q},\nu)
\Bigr )\Delta({\bf q},\nu)\ .
\end{align}
The function $\Pi({\bf q},\nu)$, Eq. (\ref{KXX}), is
calculated by extending the method employed in Refs.
\onlinecite{DEGENNES} and \onlinecite{ANI} to include the
dependence on $\nu$ and on the wave-vector ${\bf
q}$.  The calculation is valid in the {\bf  diffusive limit}, in which
$(D/2\pi T)^{1/2}$ is much larger than the mean-free path $l=v^{}_F \tau$ of the
relevant metal, where $D=l^2/(d\tau)$ is the diffusion coefficient, $v^{}_F$ is the Fermi velocity and $\tau$ is the mean free time associated with the scattering from the non-magnetic impurities. This is equivalent to the requirement that $2\pi T \tau\ll 1$. Alternatively, we require that $\tau$ be the shortest time in the problem, or that all the energies (including $\omega^{}_D$ and $T$) be smaller than $\hbar/\tau$.

Following the derivation given
 in Ref. ~\onlinecite{DEGENNES},  we
present the response function $\Pi$ in the form
\begin{align}
\Pi({\bf q},\nu)=\sum_{\omega }K({\bf q},\nu,\omega )\ .\label{KH}
\end{align}
The function $K({\bf q},\nu,\omega)$
obeys a diffusion equation (with the diffusion constant $D$) due to the scattering by the non-magnetic impurities. It also obeys a Dyson equation, which yields 
\begin{align}
K({\bf q},\nu,\omega )=\frac{2\pi N \Theta[\omega(\omega+\nu)]}{|2\omega+\nu|
+D{\bf q}^{2}}\ ,\label{HN}
\end{align}
where $N\equiv \Omega{\cal N}(0)$ is the density of states per unit energy at the Fermi energy.  The appearance of two frequencies in the denominator of Eq. (\ref{HN}) is crucial for the discussion below.
 These two frequencies result from the appearance of the two Green functions in Eq. (\ref{KXX}).
Finally, we find
\begin{align}
{\cal S}^{}_2&=\beta N\sum_{\bf
q}\sum_{\nu} a|\Delta({\bf q},\nu )|^{2}
\
,\label{ACT2}
\end{align}
where
\begin{align}
a({\bf q},\nu ,T)=\frac{1}{\lambda}-\frac{1}{\beta N}\Pi({\bf q},\nu ,T)\ .\label{ANS}
\end{align}
Since the Debye frequency $\omega^{}_{D}$ serves as a cutoff on the fermionic Matsubara frequencies  $\omega=\pi T(2n+1)$ in Eq. (\ref{HN}), one finds
\begin{align}
\frac{1}{\beta N}\Pi({\bf q},\nu ,T)=-\Psi\Bigl(\frac{1}{2} +\frac{|\nu|
+D{\bf q}^{2}}{4\pi T}\Bigr
)\nonumber\\
+\Psi\Bigl(\frac{1}{2} +\frac{|\nu| +D{\bf
q}^{2}+2\omega^{}_D}{4\pi T}\Bigr )\ ,\label{gamT}
\end{align}
where $\Psi$ is the digamma function.

Equation (\ref{ACT2}) represents the first term in an effective  GL-like expansion of the free energy density in powers of the order parameters $\Delta({\bf q},\nu)$. The original GL theory \cite{GL} was phenomenological, and the coefficient $a$ was written as $a=a'(T-T^{}_c)+c{\bf q}^2$, which is equivalent to the Ornstein-Zernike approximation. \cite {OZ} This ignored the quantum fluctuations, and was presumed valid very close to $T^{}_c$ and for long wave-lengths. An extension which includes quantum fluctuations had  $a({\bf q},\nu,T)=a'(T-T^{}_c)+c|{\bf q}|^2+e|\nu|^{m}/|{\bf q}|^{m'}$. \cite{hertz} Such forms (with $m=1$ and $m'=0$) can also be obtained from the microscopic expression (\ref{ANS}), if one expands that expression to the lowest order in $(T-T^{}_c)$, in ${\bf q}^2$ and in $\nu$. However, it is clear that an expansion of $\Psi$ in $\nu=2\pi T m$ is not justified for any $m\ne 0$, and that an expansion in powers of ${\bf q}$ may also be allowed only in a limited range of momenta. Much of the literature restricts itself to the `static' limit, i.e. $\nu=0$, and to the expansion of $a$ up to order ${\bf q}^2$. \cite{VARLAMOV}  Our purpose here is to investigate the {\bf full} expression (\ref{ANS}), and find deviations from the simpler GL theory.

Within such an effective Ginzburg-Landau theory, the phase transition
occurs when the coefficient $a({\bf q},\nu,T)$ first
vanishes as the temperature $T$ is lowered.    Using Eq. (\ref{gamT}),
this happens for the fluctuation-free ``classical", Landau or mean-field limit, ${\bf q}=\nu=0$.  For $\omega^{}_D\gg T$, we use the asymptotic limit $\Psi(z)\sim \ln z$ for large $z$ and find  the transition temperature
\begin{align}
T_c=\frac{2\gamma^{}_E}{\pi}\omega^{}_D e^{-1/\lambda}\ ,\label{Tc}
\end{align}
where $\gamma^{}_E=e^{-\Psi[1/2]}/4$ is the Euler constant. This result is consistent with the usual BCS one.

For $\omega^{}_D \gg T$ (i.e. for small $\lambda$), we replace $1/\lambda$ in Eq. (\ref{ANS}) by Eq. (\ref{Tc}). For $T$ close to $T^{}_c$, we also denote $t\equiv\ln(T/T^{}_c)$, and replace $T$ by $T^{}_c$ in the denominator of $D{\bf q}^2/(4\pi T)$. Substituting also $\nu=2\pi T m$ we find \cite{VARLAMOV}
\begin{align}
a({\bf q},\nu,T)=t+ \Psi\Bigl [\frac{1+|m|}{2}+\frac{D}{4\pi T^{}_c}{\bf q}^2 \Bigr ]-\Psi\Bigl [\frac{1}{2}\Bigr ]\ .\label{aaa}
\end{align}
This is the expression we shall use in the following calculations. In some expressions below we change notation, $a({\bf q},\nu,T)\rightarrow a({\bf q},m,T)$.

In the ``pure" Landau theory, one keeps only the ``classical" term, with $\Delta^{}_0\equiv\Delta({\bf q}=0,\nu=0)$, and one adds the quartic term in the free energy density, of order $|\Delta^{}_0|^4$, so that the Landau free energy density has the form
\begin{align}
{\cal F}^{}_{L}=-\frac{T}{\Omega}T \ln{\cal Z}^{0}=T {\cal S}\approx {\cal N}\bigl (a^{}_0|\Delta^{}_0|^2+\frac{1}{2}b|\Delta^{}_0|^4 \bigr )\ ,\label{GLF}
\end{align}
where ${\cal Z}^{0}$ contains only $\Delta^{}_0$ and (for large $\omega^{}_D$) $a^{}_0=a(0,0,T)=\lambda^{-1}-\Pi(0,0,T)/(\beta N)\approx t$. Microscopic calculations yield $b=7\zeta(3)/(8\pi^2T^2)=b^{}_0/T^2$, with $b^{}_0\cong 0.1$. \cite{VARLAMOV,com}
For metals, we write the electron energy as $E=pv/2$, and therefore the density of states (per unit volume and unit energy) at the Fermi level is given by ${\cal N}(0)=S^{}_d p_F^{d-1}/v^{}_F=S^{}_d p_F^d/(2E^{}_F)$, where $E^{}_F$ and $p^{}_F$ are the Fermi energy and momentum, $S^{}_d=A^{}_d/(2\pi)^d$, and $A^{}_d=2\pi^{d/2}/\Gamma(d/2)$ is the area of the unit sphere in $d$ dimensions ($\Gamma$ is the gamma function).
Minimizing ${\cal F}^{}_L$ with respect to $\Delta^{}_0$, one finds a non-zero $\Delta^{}_0$ below $T^{}_c$ and a jump in the specific heat (per unit volume), from zero to \begin{align}
\Delta C^{(d)}={\cal N}(0)/[bT^{}_c]\cong 10 {\cal N}(0)T^{}_c=5S^{}_d p_F^dT^{}_c/E^{}_F\ .\label{delC}
\end{align}
 However, the fluctuations at non-zero wave-vectors and frequencies give important contributions to the specific heat even above $T^{}_c$, as we discuss next.

\section{Specific heat due to fluctuations}

Substituting Eq. (\ref{ACT2}) into Eq. (\ref{Zfl}) yields a Gaussian integral, with the result
\begin{align}
{\cal Z}^{}_{\rm fl,2}\sim\prod_{\bf q}\prod_{\nu}
\frac{1}{a({\bf q},\nu ,T)}\ ,
\label{ZSOF}
\end{align}
where unimportant multiplicative factors
have been omitted.

This partition function allows the calculation of the contribution of the fluctuations to various measurable quantities. For example, the contribution to the specific heat (per unit volume) is given in $d$ dimensions by
\begin{align}
C^{(d)}_{\rm fl}=\frac{\beta^2}{\Omega}\frac{\partial^2\ln Z^{}_{\rm fl,2}}{\partial\beta^2}=-\frac{\beta^2}{\Omega}\sum_{\bf q}\sum_\nu\frac{\partial^2\ln a({\bf q},\nu,T)}{\partial\beta^2}\ .
\end{align}
 Using Eq. (\ref{aaa}), with $t=\ln(T/T^{}_c)$, we find
\begin{align}
C^{(d)}_{\rm fl}=\frac{1}{\Omega}\sum_m\sum_{\bf q}\Bigl [\frac{1}{a^2}-\frac{1}{a}\Bigr ]=\sum_m S^{}_d\Lambda^d C^{(d)}_m\ .
\end{align}
Here,
\begin{align}
C^{(d)}_m=\Lambda^{-d}\int_0^\Lambda q^{d-1}dq \Bigl [\frac{1}{a({\bf q},m,t)^2}-\frac{1}{a({\bf q},m,t)}\Bigr ]\ ,\label{CC}
\end{align}
where $\Lambda$ is the wave-length cutoff.

In contrast to the `standard' GL theory, the function $a$ in Eq. (\ref{aaa}) grows only logarithmically at large $|m|$ and $|{\bf q}|$. Therefore, both terms in Eq. (\ref{CC}) diverge unless we impose upper cutoffs on the frequencies and on the momenta.
As we discuss below, some physical results may be affected by the resulting cutoff dependence.
Equation (\ref{CC}) already contains the  cutoff $|{\bf q}|<\Lambda$. In the diffusive dirty limit, one certainly requires that $|{\bf q}|<1/l$, hence $\Lambda \lesssim 1/l$. As discussed in connection with Eq. (\ref{KXX}), we require $|\nu|<\omega^{}_D<1/\tau$, and therefore $|m|<M=\omega^{}_D/(2\pi T^{}_c)$. Since we work in a regime where $T^{}_c\ll \omega^{}_D$ [i.e. small $\lambda$, see Eq. (\ref{Tc})], $M$ is rather large. Indeed, Ref. \onlinecite{VARLAMOV} proposes using $\Lambda=1/l$ and $M=\omega^{}_D/(2\pi T^{}_c)$ (but then abandons the quantum fluctuations altogether). However, since the phonon-mediated attraction arises only for momenta within $\omega^{}_D/v^{}_F$ from the Fermi momentum, and since $l=v^{}_F\tau$ and $\omega^{}_D\ll 1/\tau$, one might argue that we should use $\Lambda=\min\{ 1/l, \omega^{}_D/v^{}_F\}=\omega^{}_D/v^{}_F$.
 \cite{PCH}

 It should be noted that in principle we has to include also fluctuations with larger wave-lengths, up to the Fermi momentum $p^{}_F$. It has been known for a long time that the short-range correlations determine the critical behavior of the internal energy. \cite{FL} In the range $\Lambda<1/l<q<p^{}_F$ i.e. on distances shorter than the mean free path, the superconducting fluctuations are those of a clean superconductor. Decreasing $l$ then yields the crossover from the clean to the dirty behavior. We are not aware of such a full calculation, and it certainly goes beyond the scope of the present paper. It is also not clear yet how to deal with the fluctuations in the intermediate range $\Lambda=\omega^{}_D/v^{}_F<q<1/l$. As we show below, our main result is not very sensitive to the choice of $\Lambda$. In any case, our calculation gives a better description of the `extremely' dirty supercunductor, when $\Lambda$ and $1/l$ approach $p^{}_F$.

 It is now convenient to switch to dimensionless quantities: ${\bf Q}= {\bf q}/\Lambda$, $\delta=\Lambda l$ and
 \begin{align}
 \gamma&=D\Lambda^2/(4\pi T^{}_c)=\delta^2/(4\pi d T^{}_c\tau)\nonumber\\
 &=M\omega^{}_D\tau/(2d)=M(\omega^{}_D/E^{}_F)p^{}_Fl/(4d)\ .\label{gamma}
 \end{align}
 In total, the problem is described by three dimensionless numbers, namely $\delta$, $\gamma$ and $M$, which depend on $\omega^{}_D/T^{}_c$, $\omega^{}_D/E^{}_F$ and $p^{}_Fl$. With these, we write
\begin{align}
a({\bf Q},m,t)\approx t+\Psi\Bigl [\frac{1+|m|}{2}+\gamma{\bf Q}^2 \Bigr ]-\Psi\Bigl [\frac{1}{2}\Bigr ]\label{aa1}
\end{align}
and
\begin{align}
C^{(d)}_m=\int_0^1 Q^{d-1}dQ \Bigl [\frac{1}{a({\bf Q},m,t)^2}-\frac{1}{a({\bf Q},m,t)}\Bigr ]\ .\label{CC1}
\end{align}
Note that $C^{(d)}_m$ depends only on $t$ and on $\gamma$.

With upper cutoffs, all the integrals over ${\bf Q}$ in Eq. (\ref{CC1}) remain finite, except for the `static' terms with $m=0$.
We therefore start with a detailed discussion of the `static' term, $C^{(d)}_0$.
 The function $a({\bf Q},0,t)$ is smallest for small $|t|$ and $|{\bf Q}|$. Defining $Q_1^2=0.001/\gamma$ (the prefactor 0.001 is arbitrary, chosen so that $\gamma{\bf Q}^2\ll 1$), we now divide the integration over ${\bf Q}$ into two regimes. In the first, for $0<|{\bf Q}|<Q^{}_1$, we use the GL-like expansion
\begin{align}
a({\bf Q},0,t)\approx t+c{\bf Q}^2\ ,
\end{align}
where
\begin{align}
c=\Psi'[1/2]\gamma\equiv \pi^2\gamma/2\ ,
\end{align}
 $\Psi'(z)$ being the derivative of $\Psi(z)$. Within this approximation, $\xi=\sqrt{c/t}/\Lambda=\sqrt{\pi \xi^{}_0 l/(8 d t)}$, is the correlation length associated with the fluctuations of the Gaussian `static' dirty mode. Here, $\xi^{}_0=v^{}_F/T^{}_c$ is the ($T=0$) coherence length of the pure superconductor, and we assume $\xi^{}_0\gg l$. \cite{BCS}

 We next write $C^{(d)}_0=C^{(d)}_{0,0}-C^{(d)}_{0,c}$, with $C^{(d)}_{0,c}=C^{(d)}_{0,1}-C^{(d)}_{0,2}+C^{(d)}_{0,3}$. Here and below, the subscript $c$ stands for `correction'. The first term in $C^{(d)}$,
\begin{align}
C^{(d)}_{0,0}=\int_0^{\infty}\frac{ Q^{d-1}dQ}{(t+c Q^2)^2}= A t^{(d-4)/2}\ ,\label{AA}
\end{align}
represents the leading singular contribution. Here, $A=I^{}_d/c^{d/2}$ and
\begin{align}
I^{}_d=\int_0^\infty x^{d-1}dx/(1+x^2)^2=(2-d)\pi\csc(d\pi/2)/4\
\end{align}
is equal to 0.5 at $d=2$ and to $\pi/4$ for $d=1,~3$. The correction terms include
\begin{align}
C^{(d)}_{0,1}=\int_{Q^{}_1}^{\infty}\frac{ Q^{d-1}dQ}{(t+c Q^2)^2}=\frac{Q_1^{d-4}}{(4-d)c^2}+{\cal O}(t)\ ,
\end{align}
\begin{align}
C^{(d)}_{0,2}=\int_{Q^{}_1}^1 \frac{Q^{d-1}dQ}{a({\bf Q},0,t)^2}= \int_{Q^{}_1}^1 \frac{Q^{d-1}dQ}{a({\bf Q},0,0)^2}+{\cal O}(t)\ ,
\end{align}
and
\begin{align}
C^{(d)}_{0,3}=C^{(d)}_{0,3'}+\int_{Q^{}_1}^1 \frac{Q^{d-1}dQ}{a({\bf Q},0,0)}+{\cal O}(t)\ ,
 \end{align}
where $C^{(d)}_{0,3'}=\int_0^{Q^{}_1} Q^{d-1}dQ/(t+c Q^2)$.
 For $d=3$, all the integrals in $C^{(3)}_{0,c}$ converge even at $t=0$. For $d\leq 2$, $C^{(d)}_{0,3'}$ diverges at $t=0$:
 $C^{(2)}_{0,3'}=\ln[c Q_1^2/t]/(2c)$ and $C^{(1)}_{0,3'}=\pi/[2(c t)^{1/2}]-1/(c Q^{}_1)$.
  This term adds a negative singular correction to $C^{(d)}_0$, causing an increase in $C^{(d)}_{0,c}$ at small $|t|$. For reasons explained below, we calculate this term for  $t=t^{}_G=t^{}_{G,static}/2^{2/(4-d)}$ [see Eqs. (\ref{tG0}) and (\ref{tG}) below].  This value of $t$ is at the border of the Ginzburg region, where we need to evaluate $C^{(d)}_{0,c}$.
 For the range of parameters used below, $C^({2})_{0,3'}$  turns out to be negligible compared to the total correction term.
 Both
 $C^{(3)}_{0,c}$ and $C^{(2)}_{0,c}$  decrease fast from large positive values  as $\gamma$ increases towards 0.5, and then decrease much more slowly  as $\gamma$ increases above 0.5. For fixed $\omega^{}_D/E^{}_F$ and $p^{}_Fl$, $\gamma$ is proportional to $M\sim\omega^{}_D/T^{}_c$ [see Eq. (\ref{gamma})], so that it increases as $T^{}_c$ decreases.

\begin{figure}[ hbtp]
\includegraphics[width=8cm]{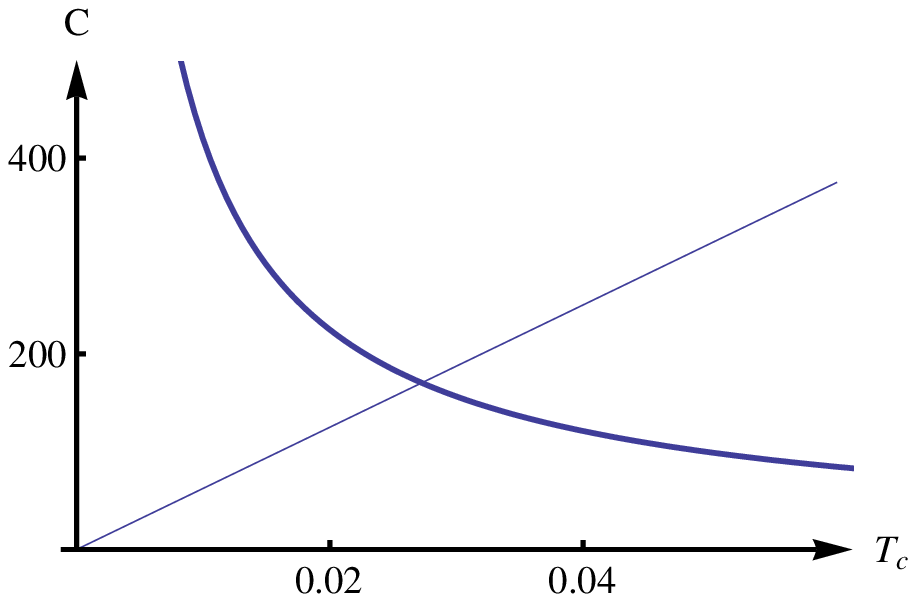}\\
\hspace{.5cm}\includegraphics[width=7cm]{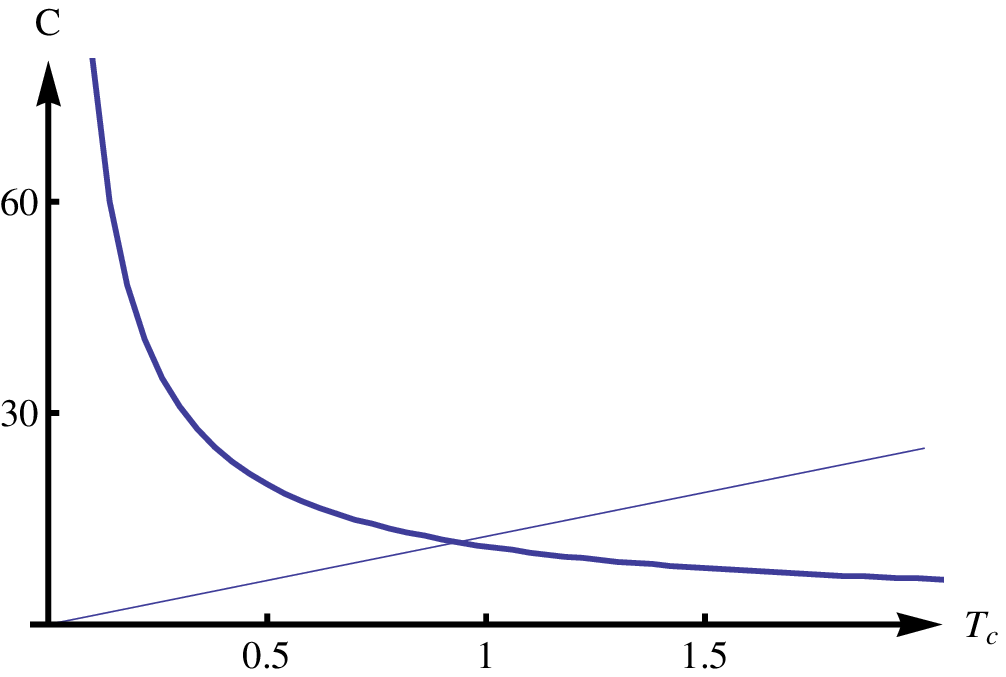}
\caption{The correction term $C^{(d)}_c$ in the Gaussian specific heat (thick line) and the discontinuity in the Landau specific heat $\Delta C^{(d)}/(S^{}_d\Lambda^d)$ (thin line) at $d=3$ (top) and at $d=2$ (bottom), for $\omega^{}_D=400$K, $E^{}_F=10^5$K and $p^{}_Fl=10$. For $C^{(2)}_c$ we used $t=t^{}_G=.4/(\pi p^{}_Fl)=0.012$.}
\label{fig1}
\end{figure}

We now turn to the quantum terms, $C^{}_m(\gamma)$ with $m\ne 0$. All these terms have finite non-zero values at $t=0$.
Furthermore, since $\Psi[(1+|m|)/2]-\Psi[1/2]>1$ for all $m\ne 0$, the integrand in $C^{}_m$ is always negative, hence $C^{(d)}_m<0$. At $\gamma=0$, one has $C^{(d)}_m=\{1/(\Psi[(1+|m|)/2]-\Psi[1/2])-1/(\Psi[(1+|m|)/2]-\Psi[1/2])^2\}/d$, which stays between $0.2/d$ and $0.25/d$ for $1\le m\le 10$  and then decreases very slowly as $m$ increases. As $\gamma$ increases, $C^{(d)}_m$ decreases slowly. In any case, the sum over $m$ grows as the cutoff $M$ grows (i.e. as $T^{}_c$ decreases), and adds to the correction term in the specific heat,
\begin{align}
C^{(d)}_c=C^{(d)}_{0,c}-2\sum_{m=1}^M C^{(d)}_m\ .
 \end{align}
 At small $T^{}_c$, $M$ is large and (unlike $C^{(d)}_{0,c}$) the total correction term $C^{(d)}_c$  increases with decreasing $T^{}_c$, see Fig. \ref{fig1}.

Writing
\begin{align}
C^{(d)}_{\rm fl}\approx S^{}_d\Lambda^d[At^{(d-4)/2}-C^{(d)}_c]\ ,\label{Ccc}
 \end{align}
 see Eq. (\ref{AA}), the correction $C^{(d)}_c$ becomes important as one moves away from $T^{}_c$ and $t$ increases.
 Equation (\ref{Ccc}) shows that the calculated $C^{(d)}_{\rm fl}$ becomes negative when $t<t^{}_0$, where
\begin{align}
t^{}_0&\equiv \Bigl (\frac{A}{C^{(d)}_c}\Bigr )^{2/(4-d)}=\Bigl [\frac{I^{}_d}{C^{(d)}_c}\Bigl (\frac{8d}{\pi}\frac{T^{}_c}{\omega^{}_D}
\frac{1}{(\omega^{}_D\tau)}\Bigl )^{d/2}\Bigr ]^{2/(4-d)}\nonumber\\
&=\Bigl [\frac{I^{}_d}{C^{(d)}_c}\Bigl (\frac{8d}{\pi}\frac{T^{}_c}{\omega^{}_D}
\frac{2E^{}_F}{\omega^{}_D}\frac{1}{(p^{}_F l)}\Bigl )^{d/2}\Bigr ]^{2/(4-d)}\ .\label{tG2}
\end{align}
This threshold decreases rather fast as $T^{}_c$ decreases. For $d\leq 2$, $C^{(d)}_c$ contains the $t-$dependent term $C^{(d)}_{0,3'}$, and one has to solve for $t^{}_0$ iteratively.
Since the specific heat should always remain positive, we conclude that the Gaussian approximation becomes questionable at $t<t^{}_0$. In that regime one should add higher order terms to the free energy. As we see below, one never reaches this `forbidden' regime, since we always have $t^{}_G \gtrsim t^{}_0$.

\section{ The Ginzburg region}
\label{Gin}

The fluctuations become dominant when their contribution to the specific heat $C^{(d)}_{\rm fl}(t)$ becomes comparable to or larger than the mean-field discontinuity $\Delta C^{(d)}$. \cite{VARLAMOV} Comparing Eq. (\ref{Ccc}) with $\Delta C^{(d)}$ gives the so-called Ginzburg criterion,
\begin{align}
t^{}_G=\bigl ([\Delta C^{(d)}/(S^{}_d\Lambda^d)+C^{(d)}_c]/A\bigr )^{-2/(4-d)}\ .\label{tG0}
\end{align}
Since we find that $C^{(d)}_c$ is mostly positive, it causes a decrease in $t^{}_G$, which becomes more and more significant as more and more quantum fluctuations are added (i.e. at lower $T^{}_c$) (although it also contains `classical' corrections from $C^{(d)}_{0,c}$). {\it Counterintuitively, quantum fluctuations may reduce the Ginzburg regime!}

Using Eq. (\ref{delC}), together with $\Lambda=\omega^{}_D/v^{}_F$, we find
\begin{align}
\frac{\Delta C^{(d)}}{S_d\Lambda^d}=5\Bigl (\frac{2E^{}_F}{\omega^{}_D}\Bigr )^d\frac{T^{}_c}{E^{}_F}\ ,\label{ddc}
\end{align}
which is linear in $T^{}_c$, with a large slope.

We now consider the dependence of $t^{}_G$ on the transition temperature $T^{}_c$. Figure 1 shows the dependence of both terms in the denominator of Eq. (\ref{tG0}),
$C^{(d)}_c$ and $\Delta C^{(d)}/(S^{}_d\Lambda^d)$, on $T^{}_c$. At high $T^{}_c$, the correction term is small, and one may neglect is and obtain the `usual' `static' Ginzburg criterion $t^{}_{G,static}$, see below. However, at low $T^{}_c$ the correction term becomes large, and it cannot be ignored. The two curves in Fig. 1 intercept at  crossover temperature $T^{}_\times$.
Although both $\omega^{}_D$ and $E^{}_F$ are much larger than $T^{}_c$,  usually  $\omega^{}_D$ is much smaller than $E^{}_F$, and therefore the slope $5(2E^{}_F/\omega^{}_D)^d$ in Eq. (\ref{ddc}) becomes quite large at $d=3$, yielding a relatively small $T^{}_\times$. For example, if we use $E^{}_F=10^5$K,  $\omega^{}_D=400$K and $p^{}_Fl=10$, the resulting value for the crossover temperature is  $T^{}_\times\approx 0.024$K (see Fig. 1).  However, the crossover temperature $T^{}_\times$ becomes significantly larger at lower dimensions: At $d=2$ we find $T^{}_\times\approx 1$K.

Since the cutoff on the momenta enters only via the dimensionless parameter $\gamma$, and since $C^{(d)}_c$ varies slowly with $\gamma$ at large $\gamma$, it turns out that the results for $T^{}_\times$ are not very sensitive to the value of the cutoff $\Lambda$. This is true for all $p^{}_Fl>1$ in $d=3$, and for $p^{}_Fl>2$ at $d=2$. Therefore, the questions raised above, before Eq. (\ref{gamma}), may not be too severe. For $d=1$ the equations yield $T^{}_c\sim 3$K, but they also yield $t^{}_G\sim 200$, which is certainly beyond the range of the approximations used above. However, the latter value is consistent with Ref. \onlinecite{HAMUTAL2}, which found very large effects of the quantum fluctuations in one dimension.

For $T^{}_c\gg T^{}_\times$ we can neglect the second term in the square brackets in Eq. (\ref{tG0}), and reproduce the `usual' static Ginzburg criterion,
\begin{align}
 t^{}_{G,static}&=\Bigl [\frac{S^{}_d I_d\Lambda^d}{c^{d/2}\Delta C^{(d)}}\Bigr ]^{2/(4-d)}\nonumber\\
 &= \Bigl [\frac{I^{}_d}{5}\Bigl (\frac{4d}{\pi}\Bigr )^{d/2}
\Bigl (\frac{T_c}{E^{}_F}\Bigr )^{(d-2)/2}\frac{1}{(p^{}_F l)^{d/2}}\Bigr ]^{2/(4-d)}  \ ,\label{tG}
\end{align}
as found in many earlier papers. \cite{PCH,VARLAMOV} Interestingly, $t^{}_{G,static}$ increases with $T^{}_c$ at $d=3$, does not depend on $T^{}_c$ at $d=2$ and deacreases with $T^{}_c$ at $d=1$. Indeed, this approximate expression has been adopted in most of the literature. \cite{VARLAMOV} However, as $T^{}_c$ approaches $T^{}_\times$ the square brackets in Eq. (\ref{tG0}) increase, and $t^{}_G$ decreases relative to Eq. (\ref{tG}). At $T^{}_c=T^{}_\times$ we have $t^{}_G=t^{}_{G,static}/2^{2/(4-d)}$, as used in the plot of $C^{(2)}_c$ in Fig. 1. Eventually, Eq. (\ref{tG}) is no longer valid for $T^{}_c<T^{}_\times$.

For $T^{}_c < T^{}_\times$, the square brackets in Eq. (\ref{tG0}) are dominated by $C^{(d)}_c$. This means that the difference between the two terms on the right hand side of Eq. (\ref{Ccc}), which is of order  $\Delta C^{(d)}$, is very small. As a result, the solution $t^{}_G$ to the equation $C^{(d)}_{\rm fl}=\Delta C^{(d)}$ becomes very close to $t^{}_0$, where the approach must be abandoned. Since we still have $t^{}_G \gtrsim T^{}_\times$, and we must stay above $t^{}_G$, we never encounter the regime $t<t^{}_0$. However, for $T^{}_c<T^{}_\times$ the higher order terms in $C^{(d)}_{\rm fl}$ may be important even above the calculated $t^{}_G$.

The `static' $t^{}_{G,static}$ and corrected (lower) $t^{}_G$ Ginzburg criteria are plotted in Fig. 2.
Unfortunately, for the parameters used above the value of $t^{}_G$ at the crossover point $T^{}_\times$ for $d=3$ becomes of order $10^{-10}$, which is  not realistic experimentally.
In contrast, at $d=2$ we find $t^{}_G\approx 0.012$, which is quite reasonable. These numbers are proportional to $1/(p^{}_Fl)^{d/(4-d)}$, so they increase with increasing disorder.

\begin{figure}[ hbtp]
\includegraphics[width=7cm]{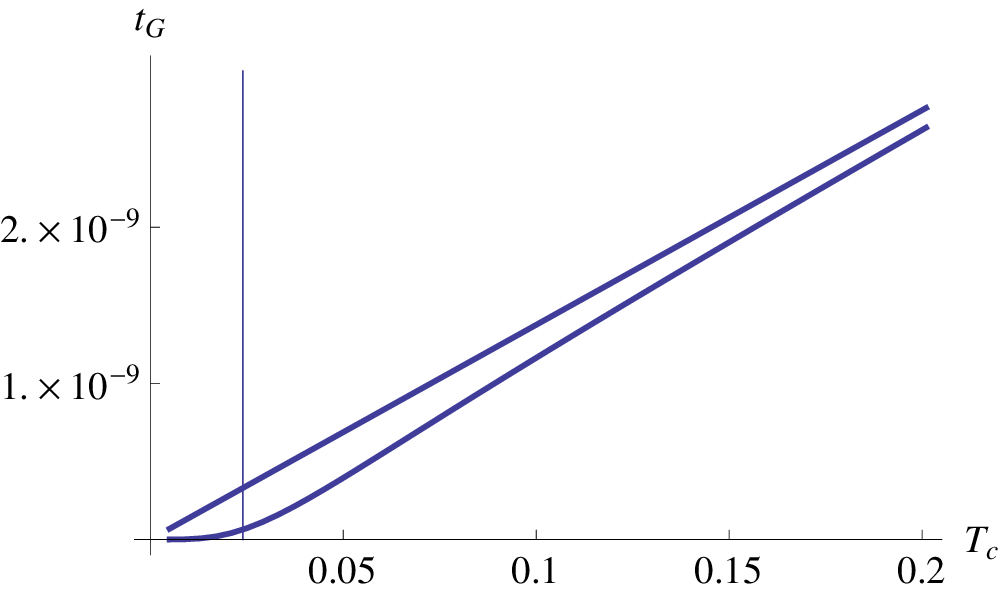}\\
\hspace{.5cm}\includegraphics[width=7cm]{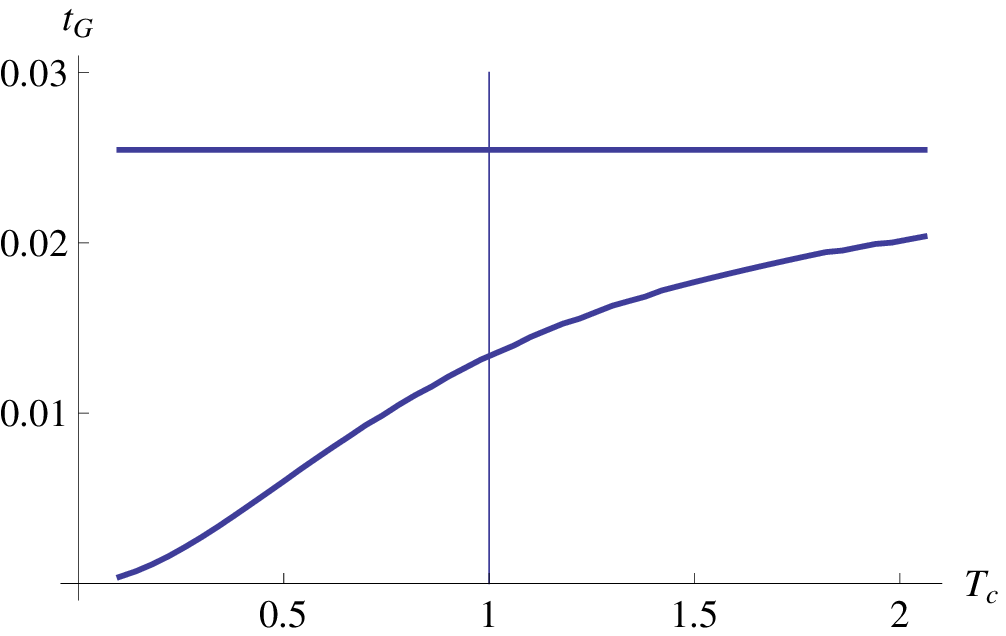}
\caption{The `static' (upper curve) and the corrected (lower curve) Ginzburg ranges versus the transition temperature, at $d=3$ (top) and $d=2$ (bottom), for the same parameters as in Fig. 1. The vertical lines show $T^{}_\times$.}
\label{fig2}
\end{figure}

\section{Summary and Discussion}

Our main result concerns the crossover temperature $T^{}_\times$, at which the (classical and quantum) corrections to the leading divergent Gaussian specific heat become equal to the Landau mean-field discontinuity in the specific heat. When the superconducting transition temperature $T^{}_c$ is larger than $T^{}_\times$, the Gaussian fluctuations (and especially the quantum ones) cause a (surprising) decrease in the Ginzburg region, by a factor which can be as large as $2^{2/(4-d)}$ [see Eq. (\ref{tG0})]. However, when $T^{}_c$ is smaller than $T^{}_\times$, the Landau mean-field discontinuity becomes smaller than the contribution from the fluctuations, so that the Ginzburg criterion implies an almost vanishing Gaussian specific heat. Therefore,  one should no longer use the specific heat to deduce the Ginzburg criterion.

In three dimensions we estimate $T^{}_\times\approx 0.025$K, which is rather low. However, $T^{}_\times$ increases at lower dimensions, which casts doubts on the use of the `standard' Ginzburg criterion for low transition temperature superconductors in those dimensions.

It should be noted that we derived the Ginzburg criterion  above $T^{}_c$, based on the specific heat. Other derivations of the Ginzburg criterion are mainly below $T^{}_c$, e.g. comparing the fluctuations in the order parameter to its average value, \cite{PCH} or comparing the ordering free energy in a coherence volume to $k^{}_BT$.\cite{FFH} In Ref. \onlinecite{we} we compared the persistent current due to the Gaussian fluctuations to that generated by the quartic term in the GL free energy.
All of these criteria give the same scaling with the basic physical parameters of the problem, but yield different prefactors. Future work should consider the effects of quantum fluctuations on these other criteria.

It should also be noted that similar correction terms arise in all phase transitions (even without the quantum fluctuations). We hope that our paper will stimulate more discussion of such corrections in other cases.

\begin{acknowledgments}
This paper is dedicated to the memory of our dear friend and colleague Yehoshua Levinson. We have collaborated with him on a variety of physics problems, and we have all learned from him a lot. He is fondly remembered.  This work was
supported by  the US-Israel Binational Science Foundation (BSF),
by the Israel Science Foundation (ISF) and by its Converging
Technologies Program. OEW and AA also acknowledge the support of the
Albert Einstein Minerva Center for Theoretical Physics, Weizmann
Institute of Science and the hospitality of the Korean Institute for Advanced Study.
\end{acknowledgments}


\begin{thebibliography}{999}
\bibitem{com}
In principle, the quartic terms in $\Delta$ may contain products of $\Delta$'s at different wave-lengths and frequencies, and the coefficient $b$ may depend on all of these different variables. 
\end{thebibliography}
\end{document}